\documentclass[%
 reprint,
nofootinbib,
 amsmath,amssymb,
 aps,
prd,
]{revtex4-2}

\usepackage{color}
\usepackage{ulem}
\usepackage[subnum]{cases}
\usepackage{slashed}
\renewcommand\sout{\bgroup \color{blue} \ULdepth=-.5ex \ULset}

\newcommand{\cL}{\mathcal{L}}
\newcommand{\pt}{\partial}
\newcommand{\lbar}{\bar{\ell}}

\newcommand{\vp}{\vec{p}}
\newcommand{\vq}{\vec{q}}
\newcommand{\vx}{\vec{x}}
\newcommand{\vy}{\vec{y}}

\usepackage{graphicx}
\usepackage{dcolumn}
\usepackage{bm}
\usepackage{hyperref}

\begin{document}

\title{Influence of dynamical screening of  four-quarks interaction on  the chiral phase diagram}

\author{Micha\l{}  Szyma\'nski}
\email{michal.szymanski@uwr.edu.pl}
\affiliation{Institute of Theoretical Physics, University of Wroc\l aw, plac  Maksa  Borna  9, PL-50204 Wroc\l aw, Poland}

\author{Pok Man Lo}
\affiliation{Institute of Theoretical Physics, University of Wroc\l aw, plac  Maksa  Borna  9, PL-50204 Wroc\l aw, Poland}

\author{Krzysztof Redlich}
\affiliation{Institute of Theoretical Physics, University of Wroc\l aw, plac  Maksa Borna  9, PL-50204 Wroc\l aw, Poland}
\affiliation{Polish Academy of Sciences PAN, Podwale 75, PL-50449 Wroc\l aw, Poland}

\author{Chihiro Sasaki}
\affiliation{Institute of Theoretical Physics, University of Wroclaw,
PL-50204 Wroc\l aw, Poland}
\affiliation{International Institute for Sustainability with Knotted Chiral Meta Matter (WPI-SKCM$^2$), Hiroshima University, Higashi-Hiroshima, Hiroshima 739-8526, Japan}

\date{\today}

\begin{abstract}
We investigate the effect of  screening of the four-quarks contact  interactions by the ring diagram at finite temperature and density in an effective chiral model inspired by QCD in the Coulomb gauge. {As a consequence, a medium-dependent coupling naturally emerges which, in a class of chiral models, brings the chiral crossover temperature down to the value calculated in  LQCD at low net-baryon density. Furthermore,  it implies a stronger divergence of the chiral susceptibility at the critical point compared to the mean-field dynamics. 
{At vanishing temperature, however, the transition sets in at unphysically small chemical potential,  indicating a need for additional effects to compensate for the screening strength. } 
} 
We  discuss  the properties of  an effective potential for a class of models described by momentum-independent gap equations. {In particular,  we introduce the method to construct an approximate effective potential from the gap equations to determine the location of the first-order phase transition.
 }
\end{abstract}

\maketitle

\section{Introduction}

Understanding the phase structure of QCD at finite temperature and density is one of the important objectives of high-energy physics and numerous experimental~\cite{Braun-Munzinger:2015hba,Luo:2017faz,Andronic:2017pug,Bzdak:2019pkr} and theoretical efforts~\cite{Fukushima:2010bq,Fukushima:2013rx,Guenther:2020jwe,Ratti:2022qgf,Aarts:2023vsf} are put to achieve this goal. From the theoretical perspective, numerical methods of lattice QCD (LQCD) provide invaluable first-principle information at small chemical potential but current numerical techniques fail for larger $\mu_B$. 

In the absence of reliable first-principle methods, effective models are often used to study the large $\mu_B$ part of the QCD phase diagram. {They} also allow us to qualitatively investigate the effects of criticality and link them to phenomenological observations and LQCD results. This makes {effective models} excellent explanatory tools to investigate universal properties of QCD matter under extreme conditions of high  temperature and density.

In order to obtain reliable model predictions of the QCD phase diagram, it is important to compare model results on different observables with the available LQCD data. In addition to finite temperature, magnetic field~\cite{Kharzeev:2012ph,Shovkovy:2012zn,Andersen:2014xxa,Miransky:2015ava} provides an additional parameter in which the lattice simulations are possible. Finite magnetic field has non-trivial impact on chiral dynamics. At low temperatures, the chiral condensate increases with magnetic field (known as magnetic catalysis). At larger temperatures magnetic field enhances the melting of the condensate and, in consequence, the critical temperature decreases with the field intensity (the effect named the inverse magnetic catalysis). While most chiral models can capture the first effect, they tend to predict the opposite trend on the pseudo-critical temperature in function of $B$. In Refs.~\cite{Lo:2021pag,Lo:2021buz} we demonstrated that screening of the four-quark interaction by polarization allows to generate the inverse magnetic catalysis at finite temperature and magnetic field. It also naturally connects deconfinement temperature $T_d\approx 270\,$MeV of pure SU(3) theory with the chiral transition temperature for light quarks ($T_{pc}\approx156.5\,$MeV) without artificial  {adjustments} of model parameters.

In this work, motivated by promising results obtained for finite magnetic field, we explore the effect of screening of the four-quark contact interaction by the ring diagram at finite temperature and baryon chemical potential -- the conditions most relevant for relativistic heavy-ion collisions. Particularly, we explore the effect of screening on the chiral phase transition. We find that the screening improves the description of the QCD phase diagram at low densities. We discuss a method for constructing the effective potential for a class of models obtained from Dyson-Schwinger equations with momentum-independent interactions. We argue that this is not always possible and provide a simple criterion for the existence of the potential. We find that screening leads to the artificially small critical chemical potential at $T=0$. This indicates that screening, as implemented in this work, becomes too strong at larger densities which indicates the need for additional effects to compensate for its strength. 

The paper is organized as follows: In Sec.~\ref{sec:theory} we provide details of the current model. Section~\ref{sec:numerical_results} contains numerical results  on the quark condensate and Polyakov loop obtained within the model. In Sec.~\ref{sec:effective_potential} we discuss the construction of an effective potential from the available gap equations and present its approximation. In Sec.~\ref{sec:phase_structure} we present the phase diagram and explore its scaling properties. We also discuss how the regularization scheme dependence manifests itself in the present model. Finally, in Sec.~\ref{sec:conclusions} we conclude our work.

\section{Theoretical setup}
\label{sec:theory}
\subsection{Chiral model with screened interaction}
\label{sec:model_intro}

As a starting point for exploring the screening effects in a strongly interacting medium, we consider the following chiral Lagrangian motivated by the Coulomb gauge QCD~\cite{Govaerts:1983ft,Kocic:1985uq,Hirata:1989qp,Alkofer:1989vr,Schmidt:1995gea,Lo:2009ud,Reinhardt:2017pyr,Quandt:2018bbu}\,,
\begin{eqnarray}
    \cL(x)&&=\bar{\psi}(x)(i\gamma^\mu\pt_\mu-m_0)\psi(x) \nonumber\\ 
    &&-\frac{1}{2}\int d^4y\bar{\psi}(x)\gamma^0T^a\psi(x)V^{ab}(x-y)\bar{\psi}(y)\gamma^0T^b\psi(y)\nonumber \,,\\
    \label{eq:lagrangian}
\end{eqnarray}
where $T^a$ are the generators of the SU(N$_c$) group, $a=1,\,\dots,\, $N$_c^2-1$, and $V^{ab}(x-y)$ is the interaction potential. In this work, we consider a two-flavor, three-color system with a degenerate current quark mass $m_0$.

In case of an instantaneous and color-diagonal potential,
\begin{equation}
V^{ab}(x-y)=\delta^{ab}\delta(x_0-y_0)\times V(\vx-\vy)\,,
\end{equation}
the gap equation is known (see eg. Refs.~\cite{Lo:2009ud,Lo:2021pag,Lo:2021buz} for details). In this work we consider the contact potential, $V(\vx-\vy)=V_0\delta(\vx-\vy)$, for which the leading-order gap equation reduces to the self-consistent equation for the (momentum-independent) dressed quark mass $M$,
\begin{equation}
\label{eq:M_gap_eq}
    M=m_0+C_FV_0\int\frac{d^3q}{(2\pi)^3}\frac{M}{2E}(1-N_{th}(E,\mu)-\tilde{N}_{th}(E,\mu))\,,
\end{equation}
where $E=\sqrt{\vq^2+M^2}$, $C_F=(N_c^2-1)/(2N_c)$ arises form the quadratic Casimir operator,
\begin{equation}
    \sum\limits_{a=1}^{N_c^2-1}T^aT^a=C_F\mathcal{I}_{N_c\times N_c}\,,
\end{equation}
and
\begin{eqnarray}
        N_{th}(E,\mu)&&=\frac{1}{e^{\beta(E-\mu)}+1}\nonumber\,,\\
        &&\\
        \tilde{N}_{th}(E,\mu)&&=\frac{1}{e^{\beta(E+\mu)}+1}\nonumber\,,    
\end{eqnarray}
are Fermi-Dirac distributions for quarks and anti-quarks, respectively, with $\mu$ being the quark chemical potential. 

Equation \eqref{eq:M_gap_eq} has the same form as the gap equation of the Nambu--Jona-Lasinio (NJL) model~\cite{Nambu:1961tp,Nambu:1961fr}, (see also Refs.~\cite{Klevansky:1992qe,Buballa:2003qv} for the detailed discussion) in the mean-field approximation if one identifies~\cite{Lo:2021pag,Lo:2021buz}
\begin{equation}
\label{eq:NJL_replacement}
    C_FV_0\rightarrow 4N_cN_f(2G_{NJL})
\end{equation}
with $N_c=3$ and $N_f=2$. However, the present model is more natural as an effective model of QCD. Its vector-vector interaction resembles the structure of QCD quark-quark interaction due to the gluon exchange, both in the color and Dirac sectors. While there is no explicit scalar-scalar channel in the Lagrangian, the effective scalar-scalar interaction, responsible for the spontaneous breaking of chiral symmetry, is generated from the Fock-type exchange. Furthermore, systematic improvements of the quark potential can be incorporated by taking into account features of gluon propagators. 

As an example, we consider the extension to include the effects of the in-medium screening~\cite{Lo:2009ud} which proceeds by introducing the effective interaction,
\begin{equation}
\label{eq:pot_gap_eq}
    \tilde{V_0}^{-1}=V_0^{-1}-\frac{1}{2}N_f\Pi_{00}\,,
\end{equation}
which may be understood as the dressing of the gluon propagator by the Debye mass. In the ring diagram approximation, the polarization reads
\begin{equation}
    \Pi_{00}(p_0,\vp)=\frac{1}{\beta}\sum\limits_{n}\int\frac{d^3q}{(2\pi)^3}\textrm{Tr\,}\left(\gamma^0S(q)\gamma^0S(q+p)\right)\,,
\end{equation}
where the sum runs over the fermionic Matsubara frequencies ($\omega_n=(2n+1)\pi/\beta$) and $S$ is the full quark propagator. The factor $1/2$ in Eq.~\eqref{eq:pot_gap_eq} is due to the color structure, $\text{Tr}\,T^aT^b=\frac{1}{2}\delta_{ab}$. Since the fermion loop contains the full quark propagator, it feeds back into the quark gap equation. In our previous work, this mechanism was crucial for generating the inverse magnetic catalysis at finite magnetic fields~\cite{Lo:2021pag,Lo:2021buz}.

The screened interaction becomes momentum-dependent even for the contact interaction~\cite{Lo:2009ud} which greatly complicates the gap equations. Thus, as an additional simplification, we consider the screening in the static limit $(p_0=0,\vp\rightarrow0)$. In this limit, the vacuum contribution to the polarization vanishes and its medium-dependent part reads
\begin{eqnarray}
    \Pi_{00}(p_0,\vp\rightarrow 0)&&=-2\beta\int\frac{d^3q}{(2\pi)^3}\left[N_{th}(E,\mu)(1-N_{th}(E,\mu))\right.\nonumber\\
    &&+\left.\tilde{N}_{th}(E,\mu)(1-\tilde{N}_{th}(E,\mu))\right]\,,    
\end{eqnarray}
which reduces to the well-known result for a massless particle,
\begin{equation}
    \Pi_{00}(p_0,\vp\rightarrow 0,M=0)=-\frac{T^2}{3}-\frac{\mu^2}{\pi^2}\,.
\end{equation}

By performing the replacement described by Eq.~\eqref{eq:NJL_replacement}, we may directly compare results obtained using the current model with the ones obtained with the NJL model. Additionally, when a finite density is considered, the vector nature of interactions leads to the dressing of the chemical potential~\cite{Lo:2009ud},
\begin{equation}
\label{eq:mu_tilde}
    \tilde{\mu}=\mu+\frac{1}{2}C_F V_0\int\frac{d^3p}{(2\pi)^3}\left[N_{th}(E,\tilde{\mu})-\tilde{N}_{th}(E,\tilde{\mu})\right]\,,
\end{equation}
and thus the chemical potential in Eq.~\eqref{eq:M_gap_eq} should be replaced by $\tilde{\mu}$. With such a dressing, the gap equation of the present model would be, at the leading order, equivalent to the gap equation of the NJL model with a vector coupling $G_V=-G$. In this work, we aim to investigate the effect of screening in a minimal setting, and thus we neglect the dressing of the chemical potential. 

\subsection{Coupling to the Polyakov loop}
\label{sec:L_coupling}

\begin{figure*}[t]
\includegraphics[width=0.49\textwidth]{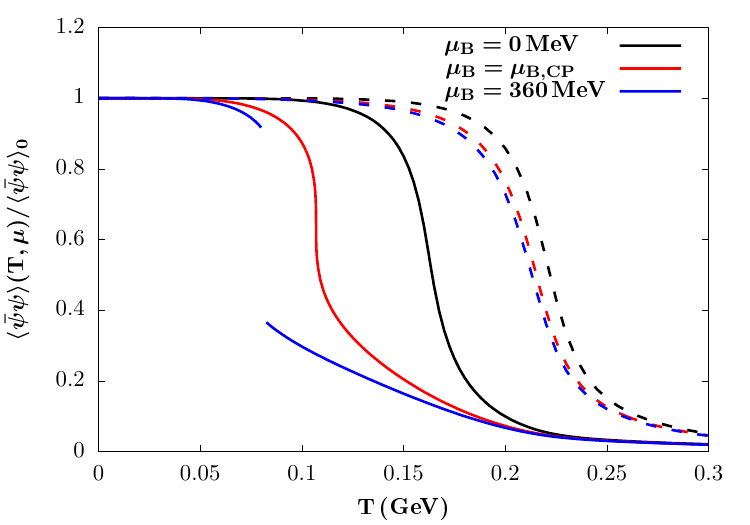}
\includegraphics[width=0.49\textwidth]{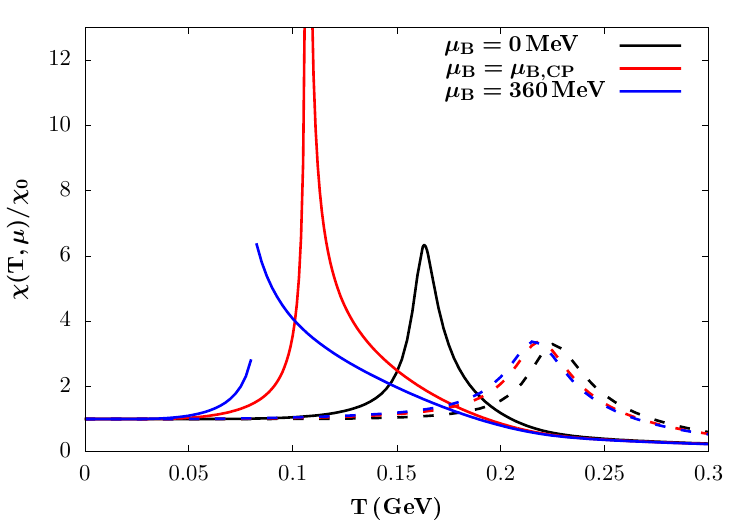}
\caption{Chiral condensate (left) and the chiral susceptibility (right) normalized to the vacuum values obtained in the current model (solid lines), compared with the PNJL model results (dashed lines), calculated for the vanishing baryon chemical potential (black), $\mu_{B,CP}$ (red) and $\mu_B=360$\,MeV (blue). \label{fig:cond_sus}}
\end{figure*}

\begin{figure*}[t]
\includegraphics[width=0.49\linewidth]{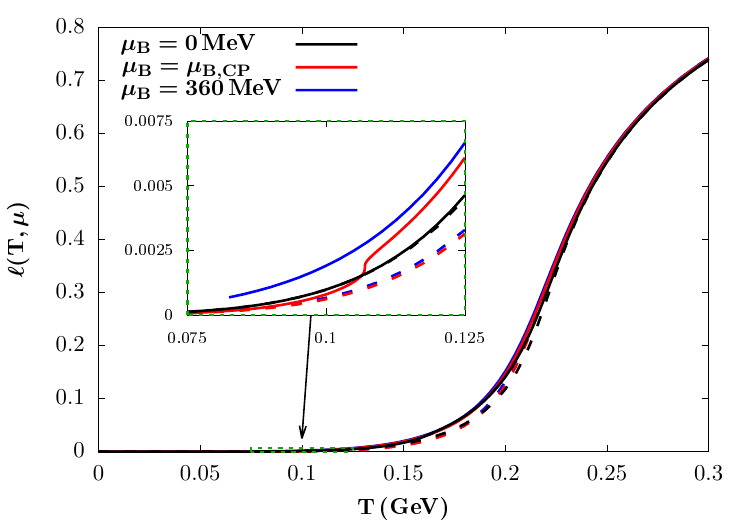}
\includegraphics[width=0.49\linewidth]{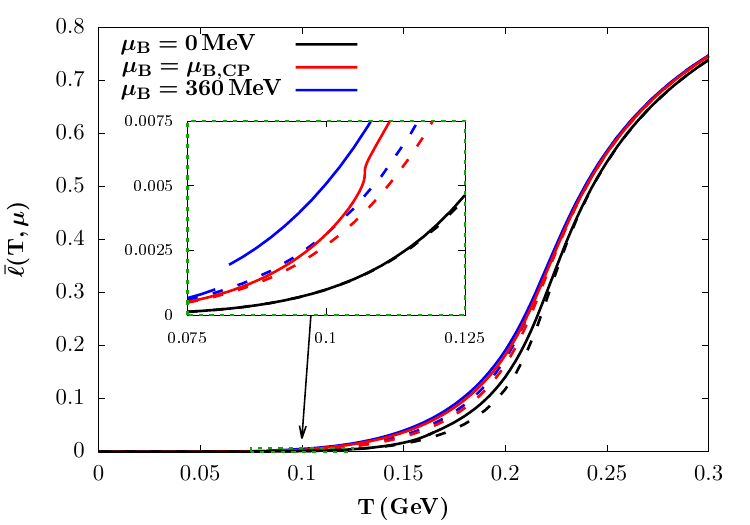}
\caption{The same as in Fig.~\ref{fig:cond_sus} but for the Polyakov loop (left) and its conjugate (right). The inset figures provide a closeup of the parts affected by the phase transition and critical point. \label{fig:l_lbar}}
\end{figure*}

To include effects of confinement which are not present in the original NJL model, we introduce the statistical confinement scheme by coupling the model to the Polyakov loop~\cite{Fukushima:2003fw,Sasaki:2006ww,Fukushima:2017csk,Lo:2013hla,Lo:2014vba,Lo:2021qkw,Kovacs:2022zcl}. This turns out to be crucial for regulating the screening strength and leads to phenomenologically correct values of the pseudo-critical temperature at the vanishing baryon chemical potential~\cite{Lo:2021buz}. 

The coupling to the Polyakov loop modifies the quark distribution function (for N$_c=3$) as follows~\cite{Hansen:2006ee},
\begin{equation}
    N_{th}(E,\mu)\rightarrow N_{th}(E,\ell,\bar{\ell},\mu)= \frac{1}{3}\sum\limits_{i=1}^{3}\frac{\hat{\ell}_F^{(i)}}{\beta(E-\mu)+\hat{\ell}_F^{(i)}}\,,
\end{equation}
where $\hat{\ell}_F$ is the Polyakov loop operator in the fundamental representation. It can be parametrized as~\cite{Lo:2021qkw}
\begin{equation}
    \hat{\ell}_F=\textrm{diag}(e^{i\gamma_1},e^{i\gamma_2},e^{-i(\gamma_1+\gamma_2)})
\end{equation}
and the Polyakov loop and its conjugate are defined as
\begin{eqnarray}
    \ell&&=\frac{1}{3}\textrm{Tr}\,\hat{\ell}_F=\frac{1}{3}\left(e^{i\gamma_1}+e^{i\gamma_2}+e^{-i(\gamma_1+\gamma_2)}\right)\,,\nonumber\\
    \bar{\ell}&&=\frac{1}{3}\textrm{Tr}\,(\hat{\ell}_F)^\dag=\frac{1}{3}\left(e^{-i\gamma_1}+e^{-i\gamma_2}+e^{i(\gamma_1+\gamma_2)}\right)\label{eq:l_lbar_definiton}\,.    
\end{eqnarray}
Using these definitions, one obtains the well-known form of the distribution function~\cite{Hansen:2006ee},
\begin{equation}
    N_{th}(E,\ell,\bar{\ell},\mu)=\frac{\ell z+2\bar{\ell}z^2+z^3}{1+3\ell z+3\bar{\ell} z^2+z^3}\,,
\end{equation}
where $z=\exp(-\beta(E-\mu))$. The corresponding distribution function for anti-particles is given by $\tilde{N}_{th}(E,\ell,\bar{\ell},\mu)=N_{th}(E,\bar{\ell},\ell,-\mu)$. 

In case of the polarization the thermal factors undergo the following replacement~\cite{Lo:2021pag},
\begin{equation}
N_{th}(1-N_{th})\rightarrow [N_{th}(1-N_{th})](E,\ell,\bar{\ell},\mu)\,,    
\end{equation}
where
\begin{equation}
[N_{th}(1-N_{th})](E,\ell,\bar{\ell},\mu)=\frac{1}{3} \sum\limits_{i=1}^3\frac{ \hat{\ell}^{(i)}e^{\beta(E-\mu)}}{(e^{\beta(E-\mu)}+\hat{\ell}^{(i)})^2}
\end{equation}
which in terms of the Polyakov loop and its conjugate can be expressed as
 \begin{eqnarray}
&&[N_{th}(1-N_{th})](E,\ell,\bar{\ell},\mu)\nonumber\\
&&=\frac{\ell z^5+4\lbar z^4+3(\ell\lbar+1)z^3+4\ell z^2+\lbar z}{\left(z^3+3\ell z^2+3\lbar z +1\right)^2}\,.    
 \end{eqnarray}
The corresponding anti-particle contribution reads $[\tilde{N}_{th}(1-\tilde{N}_{th})](E,\ell,\bar{\ell},\mu)=[N_{th}(1-N_{th})](E,\bar{\ell},\ell,-\mu)$.

\subsection{Gap equations}
By combining the formulas presented in Secs. \ref{sec:model_intro} and  \ref{sec:L_coupling} we obtain the following gap equation for the quark mass,
\begin{widetext}
\begin{equation}
\label{eq:M_gap_final}
    M=m_0+\frac{8N_c N_f G}{1-\frac{8N_c^2N_f^2}{N_c^2-1}G\Pi_{00}(M,\ell,\bar{\ell})}M\,\bigg(I_{vac}(M)-\int\frac{d^3q}{(2\pi)^3}\frac{1}{2E}\big(N_f(M,\ell,\bar{\ell},\mu)+\tilde{N}_f(M,\ell,\bar{\ell},\mu)\big)\bigg)\,,
\end{equation}    
\end{widetext}
where 
 \begin{equation}
     I_{vac}(M)=\int\frac{d^3q}{(2\pi)^3}\frac{1}{2E}\,,
 \end{equation}
is a divergent vacuum contribution that has to be regularized to obtain meaningful results. Following the previous work~\cite{Lo:2021buz} we consider the proper-time scheme,
\begin{equation}
\label{eq:schwinger_reg}
    I_{vac}(M)=\int\limits_{1/\Lambda^2}^\infty\frac{ds}{16\pi^2}\,\frac{1}{s^2}e^{-M^2s}\,,
\end{equation}
with the parameters: $G\Lambda^2=3.668$, $\Lambda=1.101\,$GeV and $m_0=5\,$MeV which give the following vacuum values of $m_{\pi}=137.8\,$MeV, $f_{\pi}=92.9\,$MeV and $\langle\bar{\psi}\psi\rangle = -(250\, \text{MeV})^3$.

To determine the expectation values of the Polyakov loop and its conjugate two additional gap equations are necessary. Here we assume their form to be
\begin{eqnarray}
    \frac{\partial }{\partial \ell}(U_G(\ell,\bar{\ell})+U_Q(M,\ell,\bar{\ell}))&&=0\nonumber \,,\\
    \label{eq:l_lbar_gap_eq}\\
    \frac{\partial }{\partial \bar{\ell}}(U_G(\ell,\bar{\ell})+U_Q(M,\ell,\bar{\ell}))&&=0 \nonumber\,,
\end{eqnarray}
where $U_G$ is the pure gauge potential and $U_Q$ describes the interaction between quarks and gluons. For the former, we choose the potential obtained in Ref.~\cite{Lo:2013hla},
\begin{eqnarray}
    \frac{U_G(\ell,\bar{\ell})}{T^4}=&&-\frac{1}{2}A(t)\ell\lbar+B(t)\ln M_H(\ell,\lbar)\nonumber\\
    &&+\frac{1}{2}C(t)(\ell^3+\lbar^3)+D(t)(\ell\lbar)^2\label{eq:gluon_potential}\,,    
\end{eqnarray}
where $t=T/T_d$ and
\begin{equation}
    M_H(\ell,\bar{\ell})=1-6\ell\bar{\ell}+4(\ell^3+\lbar^3)-3(\ell\lbar)^2
\end{equation}
is the SU(3) Haar measure, and  $T_d=270\,$MeV is the deconfinement temperature of the pure SU(3) theory. The functional form of the coefficients $A(t)$, $B(t)$, $C(t)$ and $D(t)$, as well as,  a detailed discussion can be found in Ref.~\cite{Lo:2013hla}. 

For the quark-gluon potential we employ the one-loop quark determinant in the $A_4$ background~\cite{Kashiwa:2012wa,Lo:2014vba} which in terms of Polyakov loop and its conjugate reads, 
\begin{widetext}
\begin{eqnarray}
    U_Q(M,\ell,\bar{\ell})=-2TN_f\int\frac{d^3q}{(2\pi)^3}\big[&&\ln(1+3\ell e^{-\beta(E-\mu)}+3\bar{\ell}e^{-2\beta(E-\mu)}+e^{-3\beta(E-\mu)})\nonumber\\
    +&&\ln(1+3\bar{\ell} e^{-\beta(E+\mu)}+3\ell e^{-2\beta(E+\mu)}+e^{-3\beta(E+\mu)}) \big]\label{eq:quark_potential}\,.    
\end{eqnarray}
\end{widetext}
The final set of gap equations consists of Eqs.~\eqref{eq:M_gap_final} and~\eqref{eq:l_lbar_gap_eq} which have to be solved self-consistently.

\section{Observables}
\label{sec:numerical_results}

In this section, we present the numerical results obtained using the model under consideration and compare them with the corresponding results obtained using the model without screening (which is equivalent to the PNJL model). In Figure~\ref{fig:cond_sus} (left panel) we show the quark condensate obtained from the trace of the full fermion propagator,
\begin{eqnarray}
    \langle\bar{\psi}\psi\rangle&&=-4N_cN_f M\times \nonumber \\
    &&\bigg[I_{vac}-\int\frac{d^3q}{(2\pi)^3}\frac{1}{2E}\left(N(E,\ell,\bar{\ell},\mu)+N(E,\bar{\ell},\ell,-\mu)\right)\bigg]\nonumber\,,\\
&&    
\end{eqnarray}
where $I_{vac}$ is given by Eq.~\eqref{eq:schwinger_reg}.  The chiral susceptibility, defined as
\begin{equation}
    \chi=\frac{\partial \langle \bar{\psi}\psi\rangle}{\partial m_0}\,,
\end{equation}
is shown in  Figure~\ref{fig:cond_sus} (right panel). Both quantities were normalized to their vacuum values. Solid lines correspond to the present model and dashed ones to the model without screening (which is equivalent to the PNJL model).

We first discuss the case of the present model. At the vanishing chemical potential (black) we find a substantial reduction of the pseudo-critical temperature from $T\approx230\,$MeV of the PNJL model to $T_{pc}\approx165\,$MeV. We note that such a reduction is only due to the screening and no additional tuning of the $T_d$ parameter of the pure gauge potential \eqref{eq:gluon_potential} or other modifications were necessary (see also the discussion in Refs.~\cite{Lo:2021pag,Lo:2021buz}). When the chemical potential is increased, quark condensate becomes steeper until the critical point (CP) is reached (red) which we find at $\mu_{B,CP}=314.88\,$MeV and $T_{CP}=106.95\,$MeV. For larger chemical potential we find a first-order phase transition (see the blue line for $\mu_B=360\,$MeV). On the other hand, the PNJL model shows much weaker $\mu_B$-dependence of the quark condensate, which changes only slightly for the considered range of chemical potential.  

The chiral susceptibility behaves accordingly --  for the current model, it strongly increases towards the critical point where it diverges. For larger chemical potential chiral susceptibility is finite but discontinuous which is expected behavior for the first-order phase transition~\cite{Binder_1987}. For the PNJL model, the $\mu_B$-dependence is much weaker. In fact, for the proper-time regularization scheme used in this work, the PNJL model shows no first-order phase transition in the whole $T$, $\mu_B$ plane.

In Fig.~\ref{fig:l_lbar} we show the Polyakov loop and its conjugate (the left and right panels, respectively, where the line colours and dashing are the same as in the previous figure). While these quantities are the same at the vanishing density, they become different at finite $\mu_B$~\cite{Sasaki:2006ww,Fukushima:2017csk}. We find that $\ell$ and $\bar{\ell}$ are weakly affected by the screening, in contrast to the quark condensate and chiral susceptibility. This can be understood from the fact that in the current model, there is no back-reaction of the ring diagram on the Polyakov loop gap equations \eqref{eq:l_lbar_gap_eq}. The screening enters the latter only through the dressed quark mass which changes too weakly to affect the Polyakov loop sector in a considerable manner. Nevertheless, the Polyakov loop and its conjugate remain sensitive to the critical point and the first-order phase transition, as can be seen from the inset figures.

\section{Effective potential and its approximation}
\label{sec:effective_potential}

\begin{figure}[t]
\includegraphics[width=\linewidth]{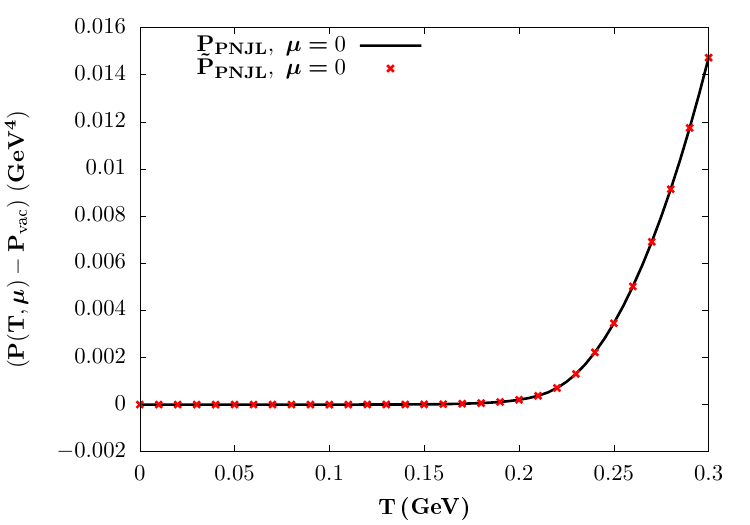}
\caption{The PNJL model pressure  for the vanishing chemical potential, obtained using the explicit mean-field formula, Eq. \eqref{eq:pnjl_potential}, (black line) and by numerical evaluation of Eq.~\eqref{eq:model_potential} (red points). \label{fig:pnjl_P_comparison}}
\end{figure}

\begin{figure}[t]
\includegraphics[width=\linewidth]{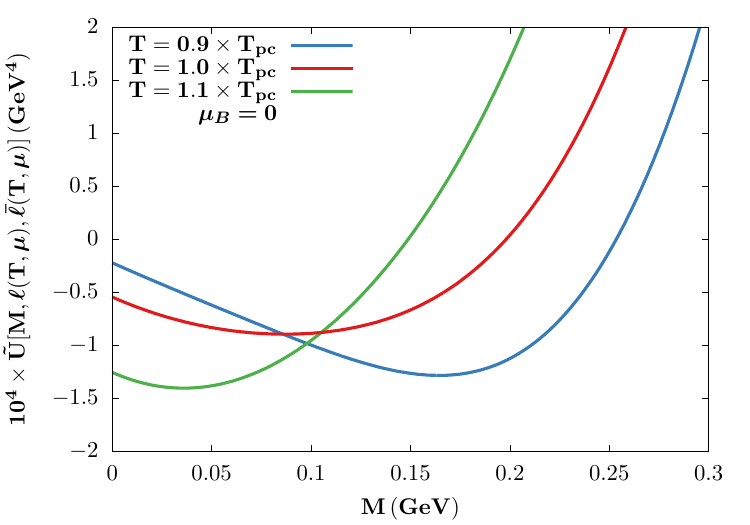}\\
\includegraphics[width=\linewidth]{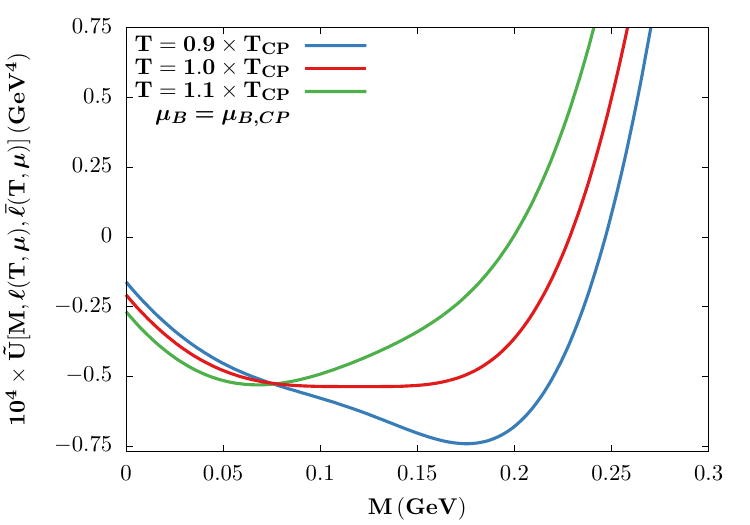}\\
\includegraphics[width=\linewidth]{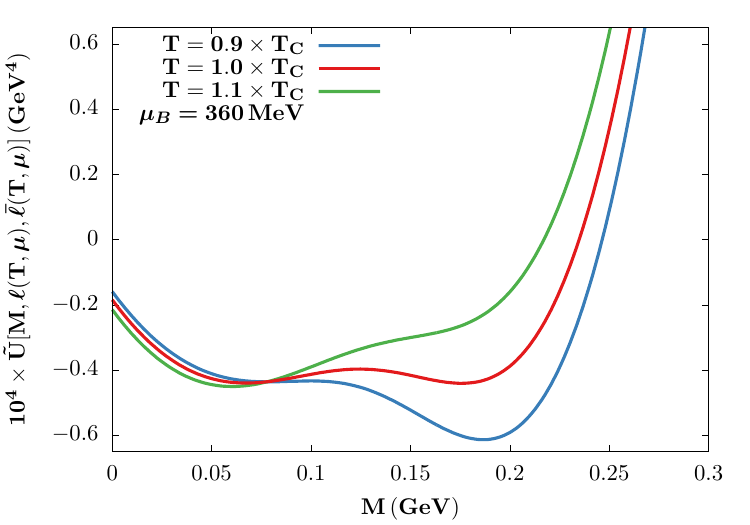}
\caption{The approximate effective potential \eqref{eq:model_potential} as a function of the dressed quark mass $M$ in case of the crossover (top panel), a critical point (middle panel) and first-order phase transition (bottom panel). See the main text for details.\label{fig:u_tilde_slices}}
\end{figure}

We find that for a sufficiently large chemical potential 
multiple solutions corresponding to local minima of the effective potential are developed, which suggests a first-order phase transition. However, the values of the order parameters obtained from the gap equations alone are not sufficient to determine which solution should be chosen. The additional information can be obtained from the effective potential -- the physical solution is the one for which it takes the lowest value.

In the mean-field approximation, an explicit form of the effective potential is constructed, commonly via the Hartree-Fock approximation. 
The gap equations are readily derived, and there is no difficulty in determining the first-order transition line. 

On the other hand, the effective potential in an interacting quantum field theory can not be computed analytically in its full form 
and one has to resort to some approximation schemes.
In this work, the gap equation for the quark mass is obtained from a certain truncation of the Dyson-Schwinger equation for the full quark propagator~\cite{Lo:2009ud} in the same spirit as a random phase approximation. In order to recover the effective potential from the Dyson-Schwinger equations, one would need to perform a functional integral -- a task which, in general, is not feasible. Additionally, on top of the truncated gap equation, a Polyakov loop coupling is implemented which was not present in the original Lagrangian \eqref{eq:lagrangian}. This further complicates the task of constructing an effective potential. In the following, we develop a scheme to construct an approximate effective potential.

In the current truncation scheme, constituent quark mass and Polyakov loop are momentum-independent, and thus the problem reduces to the construction of a scalar potential function, from which Eqs. \eqref{eq:M_gap_final} and \eqref{eq:l_lbar_gap_eq} could be derived. It is instructive to examine whether such a function exists. As an illustration, we first consider a general case of two order parameters $x$ and $y$ which are to be obtained from two gap equations (here and below we suppress the additional $T$ and $\mu_B$-dependence),
\begin{eqnarray}
        f_1(x,y)&&=0\nonumber\,,\\
        &&\label{eq:general_gap}\\
        f_2(x,y)&&=0 \,,\nonumber    
\end{eqnarray}
which we assume to be well defined in the considered range of $x$ and $y$. One wishes to construct a scalar function $\phi(x,y)$ such that 
\begin{eqnarray}
\frac{\partial \phi(x,y)}{\partial x}&&=f_1(x,y)\,,\nonumber\\
&&\label{eq:potential_conditions}\\
\frac{\partial \phi(x,y)}{\partial y}&&=f_2(x,y)\,. \nonumber
\end{eqnarray}
Since the gap equations are assumed to be derivatives of $\phi$, one may attempt to construct such a function by integration, i.e.
\begin{equation}
\label{eq:phi_formula}
    \phi(x,y)=\int\limits_{x_0}^x f_1(x',y)\, dx'+c(y)
\end{equation}
where $x_0$ is the reference value of the $x$-variable and $c(y)$ accounts for an additional $y$-dependence which may be not covered by $f_1(x,y)$. By construction, the first equation of \eqref{eq:potential_conditions} is satisfied and the second one becomes
\begin{equation}
\label{eq:phi_gap_y}
    \int\limits_{x_0}^x \frac{\partial f_1(x',y)}{\partial y}\,dx'+\frac{\partial c(y)}{\partial y}=f_2(x,y)\,.
\end{equation}
which leads to the following form of $c(y)$,
\begin{equation}
\label{eq:c_form_1}
    c(y)=\int\limits_{y_0}^y f_2(x,y')dy'-\int\limits_{y_0}^y\int\limits_{x_0}^x \frac{\partial f_1(x',y)'}{\partial y}\,dx'dy'\,,
\end{equation}
where $y_0$ is the reference value of the $y$-variable. However, it is not obvious, whether such $c(y)$ remains $x$-independent. This can be checked by differentiating both sides with respect to $x$. One finds that 
\begin{equation}
\label{eq:c_form_2}
    \frac{\partial c(y)}{\partial x}=\int\limits_{y_0}^y \frac{\partial f_2(x,y')}{\partial x}dy'-\int\limits_{y_0}^y\frac{\partial f_1(x,y')}{\partial y}\,dy'\,
\end{equation}
which vanishes only if 
\begin{equation}
\label{eq:condition_xy}
    \frac{\partial f_1(x,y)}{\partial y}=\frac{\partial f_2(x,y)}{\partial x}
\end{equation}
for all $x$ and $y$. With the aid of this condition, one finds that $c(y)$ considerably simplifies 
\begin{equation}
    c(y)=\int\limits_{y_0}^y f_2(x_0,y')\,dy'\,
\end{equation}
and is manifestly $x$-independent. One can easily  verify,  that with the above  form of $c(y)$ and the condition \eqref{eq:condition_xy}, the Eq. \eqref{eq:potential_conditions} indeed holds. The role of \eqref{eq:condition_xy} is to cancel the additional contribution to gap equations \eqref{eq:general_gap} which would arise due to the differentiation of \eqref{eq:c_form_1}. Conversely, if \eqref{eq:condition_xy} would be not satisfied then derivatives of $\phi(x,y)$ would no longer coincide with $f_1$ and $f_2$.

Condition \eqref{eq:condition_xy} can be generalized to an arbitrary number of order parameters. Writing the gap equations in a vector form, 
\begin{equation}
    \vec{F}(\vec{\varphi})=\begin{pmatrix}
        f_1(\vec{\varphi})\\
        f_2(\vec{\varphi})\\
        \vdots\\
        f_n(\vec{\varphi})
    \end{pmatrix}=0\,,
\end{equation}
where $\vec{\varphi}=(x_1,\,...,\, x_n)$ is a vector of order parameters, one wishes to find a scalar function such that $\vec{F}(\vec{\varphi})=\nabla_{\vec{\varphi}} U(\vec{\varphi})$, where $\nabla_{\vec{\varphi}}$ indicates that the gradient is taken in the order parameter space. Assuming that $F(\vec{\varphi})$ is defined on the open rectangle in $\mathbb{R}^n$, then such a function exists if
\begin{equation}
\label{eq:derivative_condition}
    \frac{\partial F_i}{\partial x_j}=\frac{\partial F_j}{\partial x_i}
\end{equation}
for $i,j=1,\,...,n$~\cite{williamson2004multivariable} which generalizes \eqref{eq:condition_xy}.

In the present case, the first gap equation can be cast into the following form,
\begin{widetext}
\begin{equation}
    f_1(M,\ell,\bar{\ell})=-\frac{M-m_0}{2G}+\frac{4N_cN_f M}{1-\frac{8N_c^2N_f^2}{N_c^2-1}G\Pi_{00}(M,\ell,\bar{\ell})}\bigg(I_{vac}-\int\frac{d^3q}{(2\pi)^3}\frac{1}{2E}\big(N_f(M,\ell,\bar{\ell},\mu)+N_f(M,\bar{\ell},\ell,-\mu)\big)\bigg)\,.
\end{equation}    
\end{widetext}
and the remaining two   {functions} {are to be} obtained from Eq.~\eqref{eq:l_lbar_gap_eq}.

One can check that the consistency conditions \eqref{eq:derivative_condition} are not satisfied, and thus there is no corresponding effective potential for the current truncation scheme. This issue can be traced back to the fact, that while we consider the effect of the ring diagram on the quark mass gap equation, we do not consider its back-reaction on the Polyakov loop sector. This suggests that the chiral and deconfinement sectors should be not treated as independent.

Although, as discussed above, there is no potential from which the set of gap equations \eqref{eq:M_gap_final} and \eqref{eq:l_lbar_gap_eq}  {can be derived} one  can still determine an {\it approximate potential} to identify the first-order chiral phase transition. We expect that the majority of the information relevant to chiral phase transition can be inferred from the $M$-direction. Indeed, while the dressed quark mass changes rapidly in the region of phase transition or crossover, the Polyakov loop and its conjugate show a little change in the same temperature range, as can be seen in Figs.~\ref{fig:cond_sus} and \ref{fig:l_lbar}. Therefore, $\ell$ and $\bar{\ell}$ will be treated as a background in which a (one-dimensional) potential is calculated. 

By analogy to Eq.~\eqref{eq:phi_formula}, we consider the following model for the effective potential,
\begin{eqnarray}
\tilde{U}(M,\,,\ell(T,\mu)\,,\bar{\ell}(T,\mu))&&=U_0(m_0,\ell(T,\mu),\bar{\ell}(T,\mu)) \nonumber\\
&&+\int\limits_{m_0}^M f_1(M',\ell(T,\mu),\bar{\ell}(T,\mu))dM'\nonumber \,.\\
&&\label{eq:model_potential}
\end{eqnarray}
The first term is analogous to $c(y)$ in Eq.~\eqref{eq:phi_formula} with $m_0$ serving as a reference point and can be seen as an effective potential for bare quarks in the Polyakov loop background,
\begin{equation}
    U_0(m_0,\ell,\bar{\ell})=U_{G}(\ell,\bar{\ell})+U_Q(m_0,\ell,\bar{\ell})
\end{equation}
with $U_{G}$ and $U_Q$ given by Eqs. \eqref{eq:gluon_potential} and \eqref{eq:quark_potential}, respectively. We stress,  that at the given $T$ and $\mu_B$, the values of the Polyakov loop and its conjugate which enter Eq.~\eqref{eq:model_potential} are obtained from the gap equations \eqref{eq:M_gap_eq} and \eqref{eq:l_lbar_gap_eq}.

It is instructive to discuss properties of such constructed approximate potential. For the case without screening, we find that the pressure~\footnote{We remind that $P=-\tilde{U}[M(T,\mu),\ell(T,\mu),\bar{\ell}(T,\mu)]$, where $M(T,\mu)$, $\ell(T,\mu)$, and $\bar{\ell}(T,\mu)$ are the solutions of the gap equations} obtained from Eq. \eqref{eq:model_potential}  has the same form (up to a constant) as the pressure obtained using the explicit functional result of the PNJL model in the mean-field approximation,
\begin{eqnarray}
U_{PNJL}(M,\ell,\bar{\ell})&&=U_G(\ell,\bar{\ell})-\frac{(M-m_0)^2}{4G}\nonumber\\
&&+4 N_cN_f\,J_{vac}(M)+U_Q(M,\ell,\bar{\ell})\,, \label{eq:pnjl_potential}
\end{eqnarray}
where $J_{vac}$ is the vacuum contribution which in the present regularization scheme reads
\begin{equation}
    J_{vac}(M)=-\int\limits_{1/\Lambda^2}^\infty \frac{ds}{32\pi^2s^3}e^{-M^2s} \,.
\end{equation}
This is illustrated in Fig.~\ref{fig:pnjl_P_comparison} for the $\mu=0$ case where the PNJL pressure, obtained from the explicit functional expression, and the numerical solution of Eq.~\eqref{eq:model_potential} indeed coincide. The fact that there is no difference between both results is not surprising, since the PNJL model gap equations are obtained from derivatives of a potential. This ensures that the condition \eqref{eq:derivative_condition} is satisfied and gap equations can be integrated back to the original potential.

For the case with screening, we check whether~\eqref{eq:model_potential} exhibits properties characteristic of the effective potential near phase transition. To this end, we study its $M$-dependence for various $T$ and $\mu$ (using the corresponding values of the Polyakov loop and its conjugate, obtained from gap equations \eqref{eq:l_lbar_gap_eq}, as the background). The results are shown in Fig.~\ref{fig:u_tilde_slices} where the upper panel corresponds to the crossover, the middle panel to the critical point and the  lower panel to the first-order phase transition. In all these panels the potentials are shown  for temperatures smaller, equal and greater than the corresponding (pseudo) critical temperatures. In the case of the crossover, we always find a single minimum that moves continuously towards $M=m_0$ with increasing temperature. This behaviour persists with increasing chemical potential until the critical point is reached - in this case, we find that potential becomes flat at the critical point, indicating divergent susceptibility. For larger chemical potentials, $\tilde{U}$ develops two distinct minima, consistently with the behaviour expected for the first-order phase transition~\cite{Binder_1987}. For $T<T_C$, the global minimum corresponds to the larger value of the constituent quark mass.  At the critical temperature (which, for the value of the chemical potential used in the figure, reads $T_C=89.65\,$MeV), the potential has the same value for both minima and for $T>T_C$ the smaller value of the dressed quark mass is energetically favourable. These  features are  essential for determining the phase diagram with the first-order phase transition  in the $(T,\mu)$-plane.

We have also tested whether the constituent quark masses obtained by solving the model gap equations are consistent with the location of the minima of $\tilde{U}$. To this end, we calculated a numerical derivative $\partial\tilde{U}/\partial M$ along the solutions of the gap equations and checked  that it is indeed close to zero within numerical accuracy.

\begin{figure}[t]
\includegraphics[width=\linewidth]{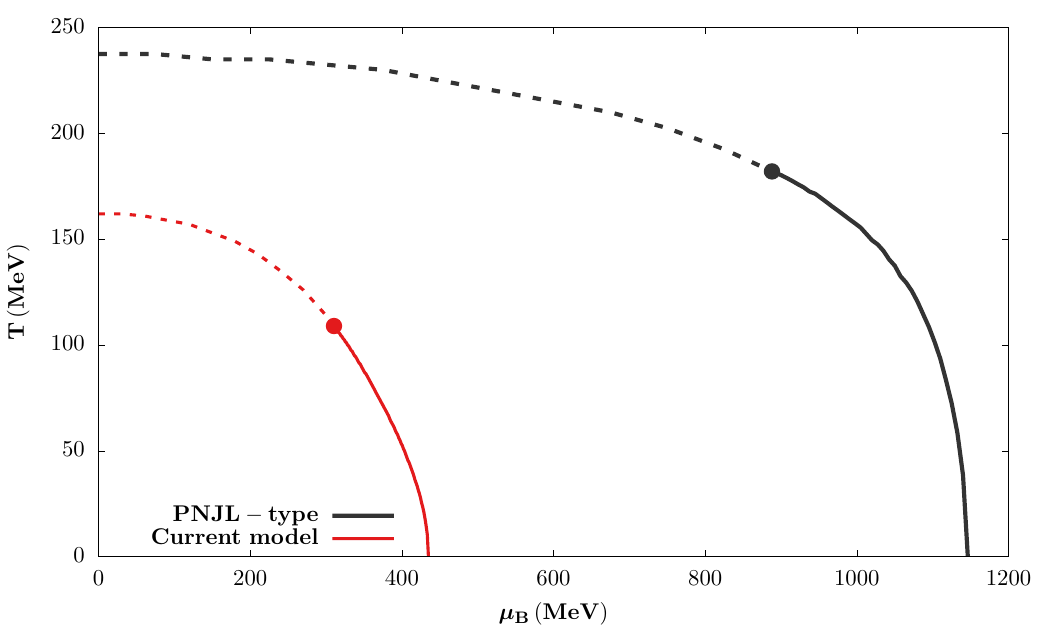}
\caption{Comparison between a typical phase diagram of the NJL model (black, obtained with 3D regularization scheme with parameter set 2 of Ref.~\cite{Buballa:2003qv}) and the current model (red, with the proper-time regularization~\ref{eq:schwinger_reg}). Solid line -- first order phase transition, dashed line -- crossover, dot -- critical point. \label{fig:phase_diagram}}
\end{figure}

\begin{figure*}[t]
\includegraphics[width=0.49\linewidth]{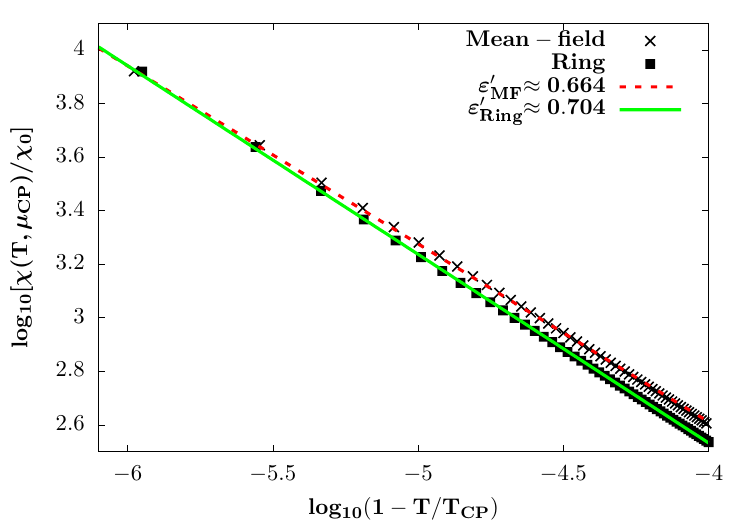}
\includegraphics[width=0.49\linewidth]{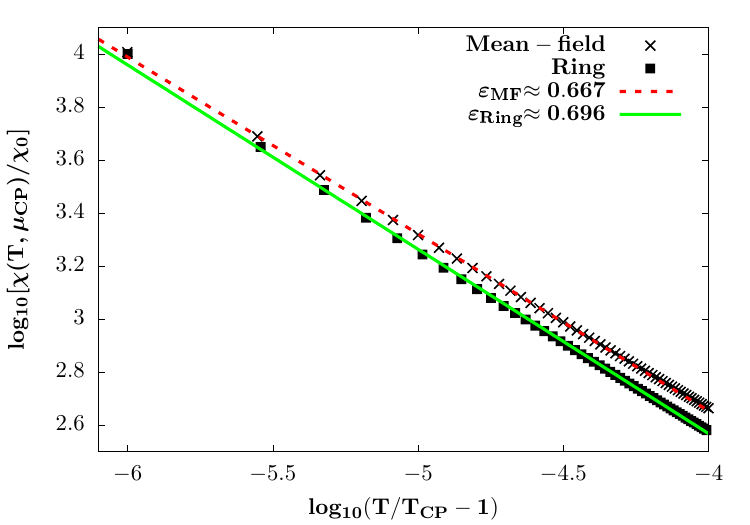}
\caption{The chiral susceptibility close to the critical point for the current model (squares) and the PNJL model (crosses) for temperature below (left) and above (right) the corresponding critical temperature at the fixed critical chemical potential. Also shown are the corresponding linear fits for the current model (solid green line) and the PNJL (red dashed line). The values of critical temperatures and chemical potentials for each model can be found in the main text.\label{fig:crit_exp}}
\end{figure*}

\section{Phase diagram}
\label{sec:phase_structure}

\subsection{Comparison with the mean-field model}
\label{sec:phase_diagram}

It is clear that Eq.~\eqref{eq:model_potential} shows the behaviour expected from the effective potential and thus it can be used to determine the first-order chiral phase transition line for the model with screened interactions. The resulting phase diagram is shown in Fig.~\ref{fig:phase_diagram} (red). For comparison, we also show a phase diagram of the PNJL model in the mean-field approximation. In this figure, dashed lines correspond to the crossover (where the pseudo-critical temperature is determined from the peak of the chiral susceptibility). Solid lines indicate  the first-order phase transition and dot  the critical point. We stress that in the case of the PNJL model, the proper-time regularization scheme \eqref{eq:schwinger_reg} does not yield the critical point in the $\mu_B$, $T$ plane. Therefore, for the PNJL model we used the 3D cutoff with the parameter set II of Ref.~\cite{Buballa:2003qv} (the regularization scheme for the model with screening remains unchanged). 

We find that screening improves the properties of the phase diagram at low baryon densities by bringing the pseudo-critical temperature closer to values  expected  in LQCD,  which are  much lower than   those obtained in  the PNJL-type of models. However, screening  leads to artificially small values of the critical chemical potential at vanishing and small temperatures. This shows that screening effects, as implemented in the current model, become too strong at larger chemical potentials. For a more accurate description of the phase diagram in high baryon density regime, a more realistic momentum-dependent interaction potential that takes into account both confinement properties and critical chiral dynamics should be considered. Additionally, other effects such as the incorporation of gluonic degrees of freedom or dressing of the chemical potential, as described by Eq.~\eqref{eq:mu_tilde} should be included. Investigation of these effects will be pursued in forthcoming  studies.

\subsection{Critical behavior}
\label{sec:criticality}

The QCD critical point is expected to belong to the $Z_2$ universality class of the three-dimensional Ising model~\cite{Wilczek:1992sf,Berges:1998rc,Halasz:1998qr,Hatta:2002sj}. Close to CP chiral susceptibility diverges with the strength depending on the direction from which the critical point is approached~\cite{Hatta:2002sj}. For paths that approach the critical point from directions that are not tangential to the first-order phase transition line, the chiral susceptibility diverges as
\begin{equation}
\label{eq:chi_scaling}
    \chi(T,\mu)\propto |g^c-g|^{-\varepsilon}\,,
\end{equation}
where  $|g^c-g|$ is the distance to the critical point in some unit~\cite{Hatta:2002sj}. In the mean-field approximation $\varepsilon_{MF}=2/3$ which is lower than the 3D Ising model universality class value $\varepsilon\approx0.78$~\cite{Schaefer:2006ds,Sasaki:2006ws}. 

By implementing the screening by the ring diagram, we go beyond the mean-field approximation and thus the value of $\varepsilon$   may change. To investigate this effect  we study a temperature dependence of the chiral susceptibility at the fixed critical chemical potential using the PNJL model in the mean-field approximation and in the current model. For the PNJL model, we use the same regularization scheme as discussed in the context of Fig.~\ref{fig:phase_diagram}. We find that the PNJL model critical point is located at $T_{CP}^{PNJL}=183.43$\,MeV and $\mu_{B,CP}^{PNJL}=883.08$\,MeV. 

The scaling results for both models are illustrated in Fig.~\ref{fig:crit_exp}, where the left panel corresponds to $T$ below $T_{CP}$ and the right panel to $T$ above $T_{CP}$. Both models show the linear behaviour in the log-log plot, consistent with the expectation from the scaling~\eqref{eq:chi_scaling}. To extract the critical exponents we performed a linear fit,
\begin{equation}
    \log_{10}(\chi(T,\mu_{CP})/\chi_0)=-\epsilon\,\log_{10}(T/T_{CP}-1)+b
\end{equation}
for $T>T_{CP}$ and 
\begin{equation}
\label{eq:fit_param}
        \log_{10}(\chi(T,\mu_{CP})/\chi_0)=-\epsilon'\,\log_{10}(1-T/T_{CP})+b'
\end{equation}
for $T<T_{CP}$. We found that $\varepsilon_{MF}=0.667$ and $\varepsilon'_{MF}=0.664$, in agreement with the expected mean-field value $2/3$. For the model with screening, we find a stronger divergence, $\varepsilon_{Ring}=0.696$ and $\varepsilon'_{Ring}=0.704$, which is closer to the 3D Ising model universality class result $\varepsilon_{}\approx0.78$. Thus, dressing by polarization not only brings the pseudo-critical temperature closer to phenomenologically expected values at low densities but also is important for a proper description of the critical properties of the system.

\subsection{Regularization scheme dependence}
\label{sec:reg_dep}

The four-fermions interaction is not renormalizable and thus the choice of the regularization scheme becomes a part of the model~\cite{Klevansky:1992qe,Buballa:2003qv}.  However, various schemes with parameters fitted to the same values of physical observables in the vacuum may give different predictions on the thermodynamic properties of a strongly interacting medium. These differences may be not only qualitative (such as the value of  the pseudo-critical temperature or the location of the critical point) but also qualitative -- for example, the existence of critical point in the $(T,\,\mu_B)$ plane is scheme-dependent in the NJL model~\cite{Kohyama:2015hix}, as well as in its Polyakov-loop extended version (for example, for the proper-time and 3-momentum cutoff schemes considered in this work). This certainly limits the predictive power of the PNJL type models.

\begin{figure}[t]
\includegraphics[width=\linewidth]{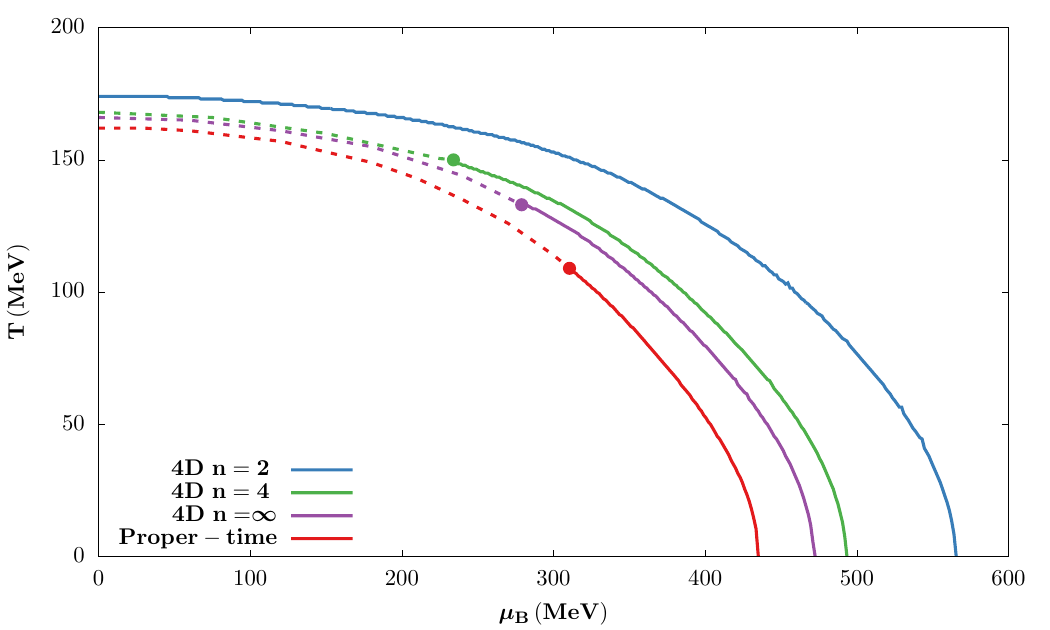}
\caption{Regularization scheme dependence of the phase diagram of the current model. Solid line -- first order phase transition, dashed line -- crossover, dot -- critical point. For the details of the regularization procedures see Sec.~\ref{sec:reg_dep}. \label{fig:reg_dep}}
\end{figure}

We find the scheme dependence of thermodynamic properties also in the present model. This is evident from Fig.~\ref{fig:reg_dep}, where phase diagrams obtained under different regularization schemes are shown. Here, apart from the proper-time regularization~\eqref{eq:schwinger_reg}, we also considered four-momentum scheme with the exponential regulator, 
\begin{equation}
    I_{vac}(M)=\int\frac{d^4q}{(2\pi)^4}\frac{\mathcal{R}_{n}}{q^2+M^2}\,,
\end{equation}
where $\mathcal{R}_{n}=exp(-(|q|/\Lambda)^n)$~\footnote{The $n\rightarrow\infty$ limit corresponds to the sharp Euclidean 4-momentum cutoff}. All the parameters were fixed to the same vacuum values of the quark condensate, pion mass and its decay constant. It can be seen that the model phase diagram depends strongly on the choice of the regularization scheme. In particular, for some regularization schemes, the transition is first-order even at the vanishing $\mu_B$. We also note that while the specific values of the (pseudo) critical temperature and chemical potential obtained in the current model depend on the choice of the regularization scheme, we find that they are smaller than the ones obtained in the PNJL model.

\section{Conclusions}
\label{sec:conclusions}

We examined the role of the screening of the four-point quark interactions in an effective chiral model at finite temperature and density. We find that the screening considerably reduces the pseudo-critical temperature at small baryon chemical potential, bringing it closer to the phenomenologically expected values. Notably, no additional modification of the model parameters was necessary to achieve this goal. The screening of the four-point quark interactions also improves the critical properties of the model by pushing  the critical exponents from their mean field towards the quantum values.  

The gap equations considered in this work originate from a set of truncated Dyson-Schwinger equations that go beyond the conventional Hartree-Fock scheme. Consequently, the explicit functional form of the corresponding effective potential is not known. We analyze the construction of this potential from the gap equations and establish a transparent criterion for its existence.

The constructed effective potential is limited by the truncation error and can not yet fully derive the given gap equations. This issue is primarily linked to the omission of screening effects on the Polyakov loop sector. Nonetheless, we successfully developed an approximate one-dimensional expression for the potential, enabling us to unequivocally identify the first-order phase transition.

We find the critical value of the baryon chemical potential at vanishing temperature to be unphysically small, indicating that screening becomes too strong at lower temperatures and larger densities. This calls for investigating the role of other potential effects, such as dressing of the chemical potential, as well as, including higher-order corrections to polarization or the vertex corrections which were not considered in the current model. On the other hand, the contact interaction suffers from the regularization scheme dependence which limits the predictive power of the model and calls for implementing more realistic, momentum-dependent, interactions. Ultimately, the role of gluon degrees of freedom, which in the current work were modelled with the Polyakov loop, should be also reexamined. Some of these interesting points will be pursued in our future studies.

\begin{acknowledgments}
M. S. acknowledges the financial support of the Polish National Science Center  (NCN) under the Preludium grant 2020/37/N/ST2/00367.
C.S.  acknowledges the support of the World Premier International Research Center Initiative (WPI) under MEXT, Japan.
We  also acknowledge the support from the Polish National Science Center (NCN) under Opus grant no. 2022/45/B/ST2/01527 (P.M.L., K.R. and C.S.). K.R. also acknowledges the support of the Polish Ministry of Science and Higher Education.
\end{acknowledgments}

\end{document}